\documentclass[letterpaper,twocolumn,10pt]{article}
\usepackage{usenix2019_v3}

\usepackage{tikz}
\usepackage{amsmath}
\usepackage{amsfonts}
\usepackage{capt-of}
\usepackage{listings}
\usepackage{makecell}
\usepackage{dcolumn}
\usepackage{tikz}
\usetikzlibrary{calc,shapes.multipart,chains,arrows,scopes,arrows.meta}
\usepackage{adjustbox}
\usepackage{xfp} %
\usepackage{siunitx}
\usepackage{ifthen}

\begin{document}

\date{}

\title{A Context-Sensitive, Outlier-Based Static Analysis to Find Kernel Race Conditions}

\author{
{\rm Niels Dossche}\\
Ghent University
\and
{\rm Bert Abrath}\\
Ghent University
\and
{\rm Bart Coppens}\\
Ghent University
} %

\maketitle

\newcommand*\halfcirc[1][1ex]{%
\trimbox{-0.5ex 0.5ex 0 0}{%
  \begin{tikzpicture}
  \draw[fill] (0,0)-- (90:#1) arc (90:270:#1) -- cycle ;
  \draw (0,0) circle (#1);
  \end{tikzpicture}}
}
\newcommand*\threefourthscirc[1][1ex]{%
\trimbox{-0.5ex 0.5ex 0 0}{%
\begin{tikzpicture}
  \draw[fill] (0,0)-- (90:#1) arc (90:360:#1) -- cycle ;
  \draw (0,0) circle (#1);
  \end{tikzpicture}
}}
\newcommand*\fullcirc[1][1ex]{
  \trimbox{0 0.5ex 0 0}{%
  \begin{tikzpicture}
    \fill (0,0) circle (#1);
    \draw[fill] (0,0)-- (0:#1) arc (0:360:#1) -- cycle ;
  \end{tikzpicture}
}}

\newcommand*\halfcircnoshift[1][1ex]{%
  \begin{tikzpicture}
  \draw[fill] (0,0)-- (90:#1) arc (90:270:#1) -- cycle ;
  \draw (0,0) circle (#1);
  \end{tikzpicture}
}
\newcommand*\threefourthscircnoshift[1][1ex]{%
  \begin{tikzpicture}
  \draw[fill] (0,0)-- (90:#1) arc (90:360:#1) -- cycle ;
  \draw (0,0) circle (#1);
  \end{tikzpicture}
}
\newcommand*\fullcircnoshift[1][1ex]{
  \begin{tikzpicture}
  \fill (0,0) circle (#1);
  \draw[fill] (0,0)-- (0:#1) arc (0:360:#1) -- cycle ;
  \end{tikzpicture}
}

\setlength{\textfloatsep}{1mm}

\newcounter{bccnt}
\newcommand{\bc}[1]{\refstepcounter{bccnt}
        \textcolor{magenta}{\textbf{Bart [\thebccnt]:} #1}}
\newcounter{bacnt}
\newcommand{\ba}[1]{\refstepcounter{bacnt}
        \textcolor{blue}{\textbf{Bert [\thebacnt]:} #1}}
\newcounter{ndcnt}
\definecolor{carrotorange}{rgb}{0.93, 0.57, 0.13}
\newcommand{\nd}[1]{\refstepcounter{ndcnt}
	\textcolor{carrotorange}{\textbf{Niels [\thendcnt]:} #1}}

\newcommand{\halo}{LLIF}

\newcommand{\roundingtwodecimalplaces}[1]{\fpeval{round(#1, 2)}\ifthenelse{\equal{\fpeval{ (floor(100*#1) == 100*#1) ? 0 : 9 }}{0}}{.00}{}}

\newcommand{\whichrate}[2]{
\csname #1#2\endcsname
}

\newcommand{\fpr}[1]{\roundingtwodecimalplaces{100 * \whichrate{#1}{falsepositives} / (\whichrate{#1}{truepositives} + \whichrate{#1}{falsepositives})}\%}

\newcommand{\totalpotentialissues}{1214}
\newcommand{\confirmedbugcount}{24}
\newcommand{\fixedbugcount}{23}

\newcommand{\modulesandcorenoheuristicstotalissues}{1107}
\newcommand{\modulesandcorenoheuristicstruepositives}{211}
\newcommand{\modulesandcorenoheuristicsfalsepositives}{648}
\newcommand{\modulesandcorenoheuristicsunknown}{248}
\newcommand{\modulesandcorenoheuristicsfpr}{\fpr{modulesandcorenoheuristics}}

\newcommand{\modulesandcoreallheuristicstotalissues}{611}
\newcommand{\modulesandcoreallheuristicstruepositives}{257}
\newcommand{\modulesandcoreallheuristicsfalsepositives}{169}
\newcommand{\modulesandcoreallheuristicsunknown}{185}
\newcommand{\modulesandcoreallheuristicsfpr}{\fpr{modulesandcoreallheuristics}}

\newcommand{\reducefprbyacceptingcomments}{0.10\%}

\newcommand{\runtimenomodulesnoheuristicssinglethreaded}{1 minute and 3 seconds}
\newcommand{\runtimenomodulesallheuristicssinglethreaded}{56 seconds}
\newcommand{\runtimenomodulesallheuristicstwothreads}{49 seconds}

\newcommand{\runtimemodulesallheuristicssinglethreaded}{8 minutes and 59 seconds}
\newcommand{\runtimemodulesallheuristicsfourthreads}{5 minutes and 42 seconds}

\definecolor{mygreen}{rgb}{0,0.6,0}
\definecolor{myred}{rgb}{0.8,0.2,0}
\definecolor{mygray}{rgb}{0.5,0.5,0.5}
\definecolor{mymauve}{rgb}{0.58,0,0.82}
  \lstdefinelanguage{diff}{
        language=C,
        morecomment=[f][\color{mygray}]{@@},
        morecomment=[f][\color{mygreen}]{+\     },
        morecomment=[f][\color{mygreen}]{+unlock},
        morecomment=[f][\color{myred}]{-\       },
}
\lstset{ %
        backgroundcolor=\color{white},   %
        basicstyle=\footnotesize,        %
        breaklines=true,                 %
        captionpos=b,                    %
        commentstyle=\color{mygreen},    %
        keywordstyle=\color{blue},       %
        stringstyle=\color{mymauve},     %
        numbers=left,
        xleftmargin=2em,
        stepnumber=1,
        basicstyle = \footnotesize\ttfamily,
        aboveskip={1.5\baselineskip},
        columns=fixed,
}

\begin{abstract}
  Race conditions are a class of bugs in software where concurrent accesses to shared resources are not protected from each other. Consequences of race conditions include privilege escalation, denial of service, and memory corruption which can potentially lead to arbitrary code execution. However, in large code bases the exact rules as to which fields should be accessed under which locks are not always clear. We propose a novel static technique that infers rules for how field accesses should be locked, and then checks the code against these rules. Traditional static analysers for detecting race conditions are based on lockset analysis. Instead, we propose an outlier-based technique enhanced with a context-sensitive mechanism that scales well. We have implemented this analysis in \halo{}, and evaluated it to find incorrectly protected field accesses in Linux v5.14.11. We thoroughly evaluate its ability to find race conditions, and study the causes for false positive reports. In addition, we reported a subset of the issues and submitted patches. The maintainers confirmed \confirmedbugcount{} bugs.

\end{abstract}

\section{Introduction}
\label{sec:introduction}

Code bases grow ever larger and more complex. Concurrency makes software even more complex and harder to understand, and comes with its own class of bugs: race conditions. A race condition occurs when a shared resource is accessed concurrently, potentially corrupting its state. While corrupted state might affect user experience and the stability of the system, many race conditions also have a security impact. They can lead to privilege escalation~\cite{dirtycow}, denial of service~\cite{cve20168655}, and memory corruption~\cite{alsarace,cve201715649}. An OS kernel is an excellent example of a complex, security-sensitive code base with lots of concurrency. A race condition in a kernel can thus be detrimental to the security of the trusted code base. This is especially problematic for cloud vendors who provide resources on shared hardware (PaaS)~\cite{lin2018measurement,manu2016study}.

Tooling is essential to help developers find race conditions, for which we consider two large classes as the root cause: the absence of locking, and the incorrect usage of locking. Both classes are made more likely by a lack of clear documentation about which locks are required to access which resources. In such cases, the developers need to rely on their own experience and on related code to judge which locks need to be applied to which fields. We call these implicit locking rules. These stand in contrast to explicit locking rules, which are documented in the code either via comments, annotations, or assertions. As most locking rules are implicit, writing concurrent code is complex and error-prone.

While dynamic analyses can be effective in helping developers detect and fix race conditions~\cite{razzer,conzzer}, and even to document locking rules~\cite{lockdoc}, static analyses have their own advantages. Not having to execute the code means even hardware-specific code (e.g., drivers) can be analysed without needing access to that hardware. Additionally, the analysis results are independent of what thread interleavings happen to occur during execution, are independent of how difficult it is to steer execution in the kernel to vulnerable code paths, and are independent of specific interactions with hardware devices. Nevertheless, it is hard to scale static analysis over a large code base and still retain adequate precision to be of help. There are many complexities in locking code that have to be accounted for, and that require additional precision: not every field in a struct might be covered by the same lock, there might be several locks covering (possibly overlapping) subsets of fields in a struct, and the racing accesses might not even be connected through the CFG making data-flow analysis more difficult, e.g., when they originate from different system calls, or callbacks.

In this paper, we present an outlier-based, context-sensitive and field-based static analysis to detect potential race conditions in complex software such as OS kernels. Our analyser statically infers locking rules from the existing code base, and subsequently tests the code base against the inferred rules to detect potential violations. The analyser only reports those whose impact passes a user-configurable threshold, in addition to having heuristics to filter certain classes of false positives. The context-sensitive mechanism enhances the outlier-based analysis' ability to prune false positive results.

In this paper, we present the following contributions:
\begin{itemize}
    \item \textbf{New context-sensitive technique:} We present a new technique that can infer field-based locking rules in complex code bases, and perform scalable outlier-based analysis based on those inferred rules, where we take the \emph{context} of similar accesses into account.%
    \item \textbf{Thorough evaluation:} We have thoroughly evaluated the results of our tool, both in finding pre-existing race conditions with a known security impact, and in finding new issues. We have categorised \totalpotentialissues{} such potential new issues in three classes: probable true positive, probable false positive, and unknown; and investigated the causes of the false positive results.%
    \item \textbf{A new open-source tool for locking bug detection:} We will release \halo{} as an open-source tool. This will allow developers of system software to check their code against our tool and possibly validate their locking rule beliefs with the inferred locking rules of our tool.
    \item \textbf{Bug fixes in Linux:} We have reported new, previously unknown bugs in Linux 5.14.11; \confirmedbugcount{} of which have been confirmed by the Linux developers. \fixedbugcount{} of those have our or the maintainer's fixes committed.
\end{itemize}

\section{Motivation and Challenges}
\label{sec:motivation}
Previous work~\cite{lockdoc,lu2008learning} shows that data races are a significant problem in complex codebases, especially when the rules about which lock protects which fields are not well-documented. There is hence a need for analysers that can automatically derive these rules and check code against them.
Static techniques typically do not know which paths can execute concurrently with other paths, which results in a high false positive rate~\cite{racerx}. To mitigate this, some tools limit themselves to very specific classes of race condition instead of general race conditions~\cite{dcuaf}. Other tools check for path feasibility, which can typically scale well for a couple of drivers, but does not scale well enough to analyse the whole kernel~\cite{deligiannis2015fast,goblint,relay}. We want a general approach that can analyse the whole kernel in a scalable way, and that can not only find bugs in drivers but also in the core kernel components.

We identify the following three challenges for static general race detection techniques for systems code: deriving locking rules, detecting fields, and filtering false positives.

\paragraph{Deriving locking rules}
Most approaches do not split the derivations of rules and the checking of rules, which makes checking the rules less scalable over large codebases. We propose a scalable, outlier-based, context-sensitive, and field-based approach to derive the rules about which locks protect which fields. An outlier-based approach can quickly filter many candidates with only intra-procedural checks. This means that it scales very well and is fairly easy to parallelise. Furthermore, this leads to an intuitive interpretation of locking rules for developers. We discuss our approach in Section~\ref{sec:design}.

\paragraph{Detecting fields}
Another challenge for field-based techniques is that code like the Linux kernel contains many object-oriented features implemented in C, despite C not being an object-oriented programming language. Such codebases therefore rely both on macros and on functions that implement these features with sometimes complex pointer arithmetic. This significantly complicates the detection of fields because the type information for a particular access is sometimes lost. Field-based approaches are therefore not always able to detect the involved fields.
Existing tools for systems code seem to rely on patches to the included macros and functions, or do not handle them at all. PeX~\cite{zhang2019pex}, for example, patches the \lstinline|container_of| macro, but this does not solve the problem in general. Furthermore, the authors of Multi-Level Type Analysis (MLTA)~\cite{mlta} tried to solve this issue in one of their implementations~\cite{mlta2}, but this functionality is disabled because the authors note it does not work properly. As we would like to have the least amount of modifications possible to kernel code, and not tie ourselves to a specific kernel (version) to analyse, we designed a more generalisable solution. We use a type-based representation for fields, which is similar to RacerD's syntactic path approach~\cite{blackshear2018racerd}. However, as RacerD only analyses Java code, it does not have to deal with pointer arithmetic complicating the correct detection of the involved field.

\paragraph{Filtering false positives}
Static approaches typically suffer from a high false positive rate. A well-known source of false positives of static race detection techniques is the fact that not all functions can execute concurrently. For example, functions that initialise~\cite{eraser,relay} or clean-up a data structure, will not concurrently access fields in that data structure. However, even when only considering code that can execute concurrently, this still leaves many benign and intentional data races as developers might not consider all races problematic. As we are more interested in races that are likely to be harmful, we want to filter these benign cases automatically. We introduce both a context-sensitive filtering approach and heuristics to reduce the number of false positives.

\section{Design}
\label{sec:design}
We base our technique on the assumption that rules dictating which fields need which locks are independent of the context, while whether a lock is required \emph{can} depend on the context. This follows the principle that locks should protect data, not code. Our intuition of a context revolves around having similar field accesses with the same locks on the same struct instance. Thus, we start by inferring \emph{locking rules} associated with \emph{field accesses}, where we match fields in a type-based manner. We identify a field by the combination of its parent type and its offset within that type. Then, when detecting violations of these rules, we take into account the \emph{context} in which the violation occurred. It is essential that we discover these rules without additional input from the tool's users or the software developers, because developers are not always aware of these rules. This is especially true in large systems code bases---such as the Linux kernel---where these rules are mostly undocumented~\cite{lockdoc}.%

At a high level, our approach consists of three phases. The first phase is an initialisation phase that consists of identifying the locks and determining which field accesses are covered by which lock. In the second phase, we derive the \emph{potential} locking rules: which fields need to be protected by which locks? We write \emph{potential}, as the next phase will filter away the rules that likely do not hold. In the third phase we detect violations against the discovered locking rules, filter them to reduce false positives, and report them to the tool's user. We use two techniques to reduce false positives. First, we utilise an \emph{outlier-based analysis} with a user-configurable threshold. Second, we take into account the \emph{context} in which a violation against a rule occurs. We only report a violation for a rule if the context likely requires locking. We describe each of these phases in more detail.

\subsection{Detecting Locks And Their Coverage}

\paragraph{Detecting Locks}

Our analysis needs to be aware of the different locks, the different types of locks, and of lock acquisition and release functions. These locking-related functions include different kinds of wrapper functions, which can perform additional functionality, such as acquiring multiple locks, reference count management, etc. These functions (transitively) use a small set of low-level locking primitives. As manually maintaining a complete list would be tedious and error-prone, only the low-level locking primitives need to be provided. The higher-level ones are automatically detected.
Similar in spirit to other techniques such as MLTA~\cite{mlta}, we consider the entire chain of the locking field. This means that, for example, acquiring / releasing \lstinline|A->B->lock| also means acquiring / releasing \lstinline|B->lock|, but not vice versa.

\paragraph{Computing Lock Coverage}
Our next step is to compute which locks \emph{definitely} cover which field accesses. Our goal is to allow fast checking of the absence or presence of a lock cover for an access, without being limited to an inter-procedural depth limitation. We initially compute this coverage intra-procedurally, after which we propagate the coverage information inter-procedurally using the following fixed-point algorithm. For each function $F$, we iterate over all its call sites and take the intersection $I$ of the lock covers of those call sites. The new lock cover set for $F$ is the union of the intra-procedural cover set of $F$ and $I$. By taking the intersection, we only get the locks that are definitely acquired across all paths towards the function. The algorithm continues as long as the last iteration still made modifications.

\subsection{Inferring Potential Locking Rules}

Next, we construct a set of potential locking rules. We construct these rules in a field-based manner. We
represent a field as the combination of the type containing the field (a struct or an array) and the offset of this field within its containing type. Since structs can be nested
within each other, we keep track of chains of fields.

To determine which locks cover which fields, we only consider locks that are in the same struct as the field, or that share a common ancestor struct. We motivate this by the observation that the locks required to protect a field are usually in a related struct.%

Multiple locks may be held during a field access. Our technique must thus be able to infer which locks are actually needed. Techniques based on lockset analysis do not report what locks are actually necessary, instead they compute the pairs of concurrently-executed functions that cause a race condition. As our approach does not use lockset analysis, and aims to report the actual rules, we approach this differently. The simplest strategy would be majority voting: one considers the lock that has been used most of the time to access the field, however that disregards that multiple locks may be necessary to protect a field. To overcome this, we use the following strategy: \emph{whenever a field access is protected with a lock, we consider this a potential locking rule}. This strategy automatically allows for multiple locks protecting one field. In the next phase these rules could be filtered away \emph{depending on the decision of the outlier-based aspect of the analysis}. This is done individually for all the locking rules.

One detail to take into account is that locks that appear together might accidentally introduce false positives for each other. Consider a struct that has two lock fields, $A$ and $B$. These are always locked in the same order: $A$ before $B$. Our algorithm would deduce that lock $B$ needs to be protected by lock $A$, as all accesses to this $B$ field are covered by $A$. However, these fields are locks and thus protection does not mean much: the locks do not always need to occur together nor do they need protection from each other. Our analysis thus ignores locking rules on fields that are locks.

\subsection{Detecting and Reporting Violations}

\paragraph{Outlier-Based Filtering of Rules}
Once we have a list of potential locking rules for every field, we can detect violations of
these rules and check if these violations should be reported. We loop over all field accesses, and for each such access, we check the rules associated with it, potentially involving chains of fields. If this field does not have any rules assigned to it, a violation cannot exist. If the field does have locking rules,
we check whether they are fulfilled by the field access; if so, there can be no violation either. Otherwise, we know that at least one path towards this access does not hold the lock, and thus is a rule violation. 
We perform outlier-based filtering to determine whether we are confident the rule is correct. We divide the detected violations by the total number of accesses. We then limit the number of reported issues by only considering rules whose fraction is lower than the threshold. The lower the fraction is, the more confident we are in the locking rule and that its violations are a real issue.

\paragraph{Contexts}
Whether a violation of a rule will be reported is also dependent on whether \emph{relevant, similar accesses} occur under the lock associated with that rule.
Having observed many accesses to a field \lstinline!A->F1! are protected with the lock \lstinline!A->L! strongly suggests that a different rule to protect \lstinline!A->F2! with \lstinline!A->L! may hold. Conversely, having observed that some accesses to \lstinline!A->P->Q->R! are protected under \lstinline!A->L! carries much less weight when deciding the correctness of the rule that \lstinline!A->F2! should be protected under \lstinline!A->L!.
We want to compare with \emph{similar accesses} (in this case, accesses that originate from \lstinline!A!), and consider the \emph{distance in terms of indirection} through which we arrive to the source from our field accesses.

To find and relate similar accesses, we first define the \emph{context} of a field access: \emph{from which sources}---operands of loads and stores, address computation in case a reference is passed to a function call---is a field access constructed. For each source involved in the field access, we compute the distance in terms of indirections between the source and the access. As fields are represented as chains, we repeat this for every field in the chain. For each rule, this results in a set of pairs that combines a field access and a context, each of which has an associated distance. We then further partition these sets along the following property: Was that field access performed under that lock, or was it performed unlocked (regarding the lock associated with that rule)? Finally, for each such partition involved, we compute a distance-weighted average of the involved accesses. We only report violations of a rule when the distance-weighted average associated with the pair is lower than the distance-weighted average of the accesses that are not locked. We ignore the operands passed to allocation functions, as they signify an initialisation and thus cannot be racy.%

\begin{figure}
	\begin{center}
		\begin{tikzpicture}[>=Stealth]
            \node[shape=rectangle,draw=black] (V) at (0,1) {vfs\_write source: argument 0};
            \node[shape=rectangle,draw=black,dashed] (ES) at (1.5,0) {an ext4 source};
            \node[shape=rectangle,draw=black,dashed] (EF2) at (1.5,-1) {ext4 field 2};
            \node[shape=rectangle,draw=black] (EF1) at (-1.5,-1) {ext4 field 1};
            \node[shape=rectangle,draw=black] (B1) at (1.5,2) {btrfs field 2};
            \node[shape=rectangle,draw=black] (B2) at (-1.5,2) {btrfs field 1};
            \node[shape=rectangle,draw=black] (B1_2) at (-4.5,1) {btrfs field 1};

            \path [-{>[scale=1.5]}] (EF1) edge node[left] {$\Delta = 1$} ([xshift=20.5pt]V.south west);
            \path [-{>[scale=1.5]}] (EF2) edge node[right] {$\Delta = 1$} (ES);
            \path [-{>[scale=1.5]}] (ES) edge node[right] {$\Delta = 2$} ([xshift=-20.5pt]V.south east);
            \path [-{>[scale=1.5]}] (B1) edge node[right] {$\Delta = 1$} ([xshift=-20.5pt]V.north east);
            \path [-{>[scale=1.5]}] (B2) edge node[left] {$\Delta = 3$} ([xshift=20.5pt]V.north west);
            \path [-{>[scale=1.5]}] (B1_2) edge node[below] {$\Delta = 2$} (V);
		\end{tikzpicture}
		\caption{Example of a diagram illustrating the context distance weighting. Dashed rectangles indicate an unlocked access, full rectangles indicates a locked access.}
		\label{fig:context_example}
	\end{center}%
\end{figure}
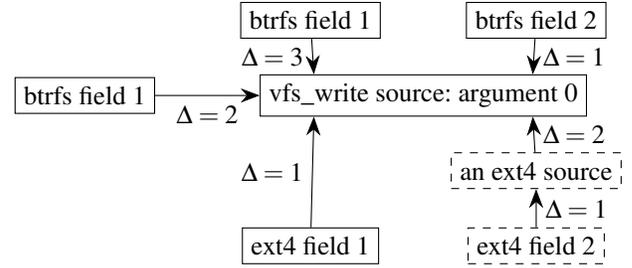
Figure~\ref{fig:context_example} serves as an example of how the distance weighting of contexts works. For clarity, we consider that there is only one possible lock involved. Distance is written along the edges with $\Delta$. Dashed rectangles indicate unlocked accesses; full rectangles indicate locked accesses. Suppose we have 2 types (ext4 and btrfs), both containing 2 fields. For all accesses to these fields, we determine the context. Suppose that these all end in argument 0 of ``vfs\_write''. There is 1 path with a distance of 1 for the ext4 type that is protected under the lock, resulting in a weighted average of 1. Similarly, there is 1 path for which the field access is not protected under the lock, which in this example has a distance of 2, resulting in a weighted average of 2. As the observations that are protected by the lock have a lower weighted average, we take this to mean that the locked accesses are more ``direct'', and we thus consider them to be more relevant. Hence, the violations of ext4's second field along this context will be reported. Applying the same computation logic for the btrfs type yields a weighted average of $\frac{2 + 3 + 1}{3} = 2$ for the locked case.

To compute the distances, \halo{} uses an inter-procedural backwards data-flow algorithm starting at every field access. The backwards data-flow pass does not stop at a single source, but collects multiple sources that are at different distances. The algorithm increases the distance only in the following situations: a load, a return value, and an address computation. The first two cases increase the distance with 1; for address computations we increase the distance with the amount of indices needed to compute the address. The intuition behind this is that a deeply nested struct is less likely to be relevant than one that is not. As mentioned earlier, we completely ignore return values of allocation functions as a possible context. The distance of an argument source in a function is further weighted according to the amount of callers that function has. The algorithm stops when a predefined depth limit is reached, or when no further backwards actions can be performed.

\section{Intentionally Unprotected Accesses}
\label{sec:heuristics}

Our technique detects unprotected accesses that might need to be protected. However, there are certain unprotected accesses that developers either know for certain to be safe, or do not care about. In this section, we present heuristics to deal with four such cases: initialisation/clean-up of structs; intentional races caused by an unlocked check--locked recheck paradigm; usage of concurrency-safe functions; and fields that are never written concurrently.

\subsection{Struct Initialisation and Clean-up}
Initialisation and clean-up code for a struct will not be executed concurrently with other accesses and will therefore be unprotected. The presented heuristics in this category are based on taint analysis. Instead of tainting field accesses, we taint the objects containing the fields, as initialisation code is written in an object-centric way. Violations against the tainted objects are filtered away as false positives.

\paragraph{Lock Initialisation}
The first heuristic that we use to detect struct initialisation is checking for lock initialisers. If a struct contains a lock, that lock must be initialised before it can ever be used. Such initialisations are only valid in the struct's initialisation function. We can detect such initialisations as non-atomic stores to memory occupied by the lock.

\paragraph{Struct Allocation or Deallocation}
When a struct has just been allocated in a function, no other references exist to that struct, and thus no violations happen during execution. Therefore, we ignore violations on newly allocated structs. Similarly, if a struct is deallocated, no references to it should exist any more, and therefore no violation can occur either. If there are still references, then a race condition would be possible in case of a violation, but that is actually a different kind of bug (e.g. a reference count bug) that can be detected by other means, and as such we decide to not consider that in this heuristic.
Allocation and deallocation function wrappers are automatically detected iteratively starting from a base list of allocation and deallocation functions. An allocation wrapper is detected when it calls another allocation function whose value is returned. A deallocation function is detected when a call to another deallocation function postdominates the first instruction after all the NULL checks.

\subsection{Unlocked Check--Locked Recheck}
Developers sometimes intentionally allow data races to speed up code. As such, these \emph{benign} data races do not affect the correctness of the program. 
We want to avoid reporting code following this paradigm to not unnecessarily burden developers.

The code example below is from the networking subsystem in Linux 5.14.11~\cite{afpacketsendmsg}:
\begin{lstlisting}[language={C++}]
if (data_race(po->tx_ring.pg_vec))
	return tpacket_snd(po, msg);
return packet_snd(sock, msg, len);
\end{lstlisting}

One of the rules our technique is able to derive from other accesses to this struct, is that \lstinline|po->tx_ring.pg_vec| needs to be accessed under the lock \lstinline|po->pg_vec_lock|, but this is not the case here. The purpose of the check is to perform an optimisation: it does not matter if the check gave the wrong result, since most of the time it will be right and \lstinline|tpacket_snd| will perform the check again anyway while executing under the expected lock. This pattern of a racy check and checking it again under a lock is quite common in Linux. Furthermore, note the use of the \lstinline|data_race| annotation, indicating that the developers know this access is racy. %

Our technique would flag this as a violation, as this check does not happen under the correct lock. To prevent this false positive case from showing up, we added support to detect this pattern. As the \lstinline|data_race| macro is only used in 100 places in the Linux kernel we evaluated our technique on, we decided not to special-case this macro, but opted for a more generic detection of this paradigm. For all detected violations, we first check if the violation occurs within a branch's compare instruction. If so, we then check whether the field is checked again under the expected lock within the branch's block. This takes into account calls to other functions. In Linux 5.14.11 we detect this paradigm at least once in 580 functions.

\subsection{Safe Functions} %
Some functions may be inherently safe, are almost always safe to use without locks, or are used where a developer might not really care that much about the raciness of the access (e.g., functions such as \lstinline|printk|, \lstinline|atomic_set|, ...). We apply a simple outlier-based heuristic to detect functions whose arguments are typically not constructed under locks. Fields that are passed to such functions will not be considered violations. 

\subsection{Write or Escape}
Many field accesses may happen under a lock by coincidence. This can be caused for example by overly large critical sections. To mitigate this, we will only consider fields that have had at least one locked write access to it, or are used as a function call argument. This gives us stronger evidence that a lock may be required for the corresponding field. This particular heuristic is also used in RELAY~\cite{relay}.

\section{Implementation}
\label{sec:implementation}
We call the implementation of our technique \emph{\halo{}}, which is short for \emph{Linux Lock Issue Finder}. Our static analysis is based on LLVM, with MLTA~\cite{crix} to increase the precision of our analysis. To increase its performance, we parallelised all steps in our analysis except for the inter-procedural fix-point algorithm to propagate the lock coverage. In the rest of this section, we discuss in more detail how we detect fields, how we count violations, and some further improvements we made to the implementation of MLTA.
\subsection{Detecting Fields}
\label{subsec:detectingfields}
For our field-based approach, we need to correctly identify which fields are accessed where. Furthermore, as mentioned earlier, we keep track of \emph{chains} of struct accesses for a certain field, i.e., an access to \lstinline|A->B->lock| implies an access to \lstinline|B->lock|, but not vice versa. Thus, it is important that we can indeed track these chains correctly. As previously stated, we would furthermore like to have the least amount of modifications possible to kernel code, and not tie ourselves to a specific kernel (version) to analyse, so we came up with a more generalisable solution to deal with the object-oriented C code in the Linux kernel. %

We perform an intra-procedural data-flow analysis starting from the field-accessing instruction's operands (i.e., address computation or bitcasts). We perform the following actions in our data-flow analysis. On a \emph{select} instruction (the LLVM equivalent to the C ternary operator) and on \emph{phi nodes}, we recurse on all the operands of the instruction. Because we also need to handle multiple layers of indirection, we also recurse over \emph{load} instructions. Finally, as Linux has structs that ``inherit'' from one another, we also recurse over \emph{(bit)casts}, while also keeping track of the different types associated with these casts.

Pointer arithmetic on address computations will be reflected in the pointer index offsets. It results in the address computation pointing outside the struct. We detect these cases and rewrite its address computation as follows. First, we compute the byte offset $B_s$ from the base pointer transitively (i.e., an address computation may have another address computation as a base pointer). We then check which bitcasts have that same base pointer as a source transitively, and have a target type $T$. We also compute its byte offset $B_c$. The original address computation can then be replaced by the address of $T$ with byte offset $B_s - B_c$ if the byte offset is non-negative. If instead there is an address computation on the bitcasted base pointer instead of just a bitcast, then we try to start our byte offset computation for $B_s$ at the address computation instead of the bitcast. This method can successfully recover the field from complex macros such as \lstinline|container_of|, and from structs that are located next to each other in memory.

If all else fails, we try to fall back to LLVM's debug metadata. However, with this debug information we can only retrieve the containing type; we cannot create full field chains due to the lack of a corresponding address computation.

\subsection{Counting Violations}
For each violation found, we need to assess how large the race condition opportunity is. Depending on the compiler's optimisation level, multiple load instructions might be merged into a single one. We want the analysis to be independent of these optimisations. We therefore weigh the value using the number of uses.%

\subsection{Improvements To MLTA}
A major issue with static analysis is the inability to precisely determine the targets of indirect calls. This is especially problematic in code bases such as kernels, as their heavy use of indirect calls hinders inter-procedural analysis. We use MLTA~\cite{mlta}, which is one of the state-of-the-art techniques to deal with this issue. In particular, we depend on Crix's implementation of MLTA~\cite{crix} to increase the precision of the ICFG and call graph\footnote{There are currently three slightly different MLTA implementations available: the one in Crix, a stand-alone implementation~\cite{mlta2}, and one integrated in another tool~\cite{mlta3}. We chose Crix's implementation as it resulted in the smallest false positive rate in our tests.}. However, we noticed that Crix's MLTA implementation unexpectedly over- and under-connected some nodes in the call graph, which we mitigated as follows.

\paragraph{Escaping Types}
MLTA falls back to function type signature matching for function pointer arguments. This is an over-approximation, and will decrease the precision of our analysis. We decided to add basic support for function pointer arguments if they are constants. In that case it is not necessary to fall back to type signature matching. 

\paragraph{Type and Function Signature Hashing}
Crix's implementation of MLTA matches types and signatures by computing IDs based on hashing the names of types and functions. However, this does not always work, as LLVM might duplicate types and give them a suffix to avoid name clashes across compilation units. This even happens with structs defined in header files. Thus, hashes for what are essentially the same types could be different, reducing the accuracy of the analysis. We have changed both MLTA's type hashing function and call / function signature hashing such that they work with a description of the fields of the structs instead of relying on the name. %

\paragraph{Address-Taken Functions}
Only address-taken functions are considered potential indirect call targets. As discussed before, the types in different compilation units might be duplicated, and as a consequence the signatures of functions in them might be different. If a compilation unit performs a direct call to another unit, the function signature might need to be casted. Therefore, the function will become address-taken and incorrectly assumed to be a potential indirect call target. This causes an over-approximation of the call graph. We have fixed this problem by detecting cases like this.

\section{Discussion}
\label{sec:discussion}
Before discussing the evaluation of our technique, we briefly discuss our work in a broader context: the documentation of expectations and rules for locking, extensions of our technique, and ethical considerations regarding this work.

\subsection{Documenting Locking Rules}
One clear issue is the lack of documentation of the rules developers should adhere to regarding accessing certain fields and data structures~\cite{lockdoc,lu2008learning}.
Better and more clearly documented rules and annotations not only help static analyses, they also aid dynamic analyses~\cite{lwn_concurrency_part2}, and humans performing manual reviews. %
Furthermore, Linux even has a \lstinline|data_race| macro that allows developers to explicitly label \emph{data races by design}~\cite{lwn_concurrency_part2}. Unfortunately, in Linux v5.14.11, there are only 100 instances of this macro in use. While this has in the meantime increased to 128 in Linux v5.19-rc1, there are undoubtly many more instances where the use of this macro could be useful. Similarly, the \lstinline|lockdep_assert_held| function documents expectations on locks. %
As with LockDoc~\cite{lockdoc}, \halo{} could aid developers in documenting rules.

\subsection{Extensions} 
\label{sec:extensions}
The information gathered by \halo{} can be used to find additional problematic code patterns. We extend \halo{}'s analyses with two such extensions: validating Lockdep assertions and finding reader-writer lock violations.

\paragraph{Statically Validating Lockdep Assertions}

Lockdep~\cite{lockdep} is a dynamic analysis tool for the Linux kernel. Its goal is to validate the rules about dependencies between different locks upon acquisition. Our analysis tool can be extended to statically check whether all paths that reach the assertions actually hold the expected lock. This coverage information is already present because we need to be able to determine the lock coverage for our regular analysis. One might be tempted to simply ignore the implied locking rules when encountering a Lockdep assertion. This intuition is wrong since there is no guarantee that the assertions are complete: it might be the case that some \emph{additional} locks are actually required.

\paragraph{Finding Reader-Writer Lock Violations}
Reader-writer locks are designed to allow multiple readers to concurrently access the same critical section, whereas writers get exclusive access to that critical section. We have extended our initial list of locking functions to indicate whether the function is meant for read locking or for write locking. Its wrappers will exhibit the same read-write semantics. The other components of \halo{} work as before, but we additionally check for the read-write semantics.

\subsection{Ethical Considerations}
There are 2 ethical considerations to be made for this paper: we find potentially security-related locking vulnerabilities; and we interact with people: the Linux kernel developers.
As for the first aspect: \halo{} reports \emph{violations} of potential locking rules, it does \emph{not} report concrete inputs that could trigger violations. %
Still, to mitigate the impact, we proactively sent fixes to 32 of the issues to the Linux kernel developers. As for the second aspect, we interact with the Linux kernel developers and report on our interactions with them (i.e., whether they accepted our patches). We got a waiver of our faculty's ethics committee/IRB for these interactions, given that we would not deceive the kernel developers. Furthermore, we adhere the Linux Foundation's Technical Advisory Board's guidelines for researchers~\cite{linux_researcher_guidelines}, which states clearly that they consent to be interacted with for receiving \emph{good faith contributions}, which our contributions are.

\section{Evaluation}
\label{sec:evaluation}

We evaluate \halo{} on Linux kernel version 5.14.11 using a default Debian server configuration on x86\_64 compiled with the default flags and \texttt{-O2 -g}.
This section is structured as follows. First, we evaluate the effectiveness of our technique on finding pre-existing race conditions with a known security impact. Then we apply \halo{} to find \emph{new} race conditions, and investigate the reported issues in detail: we investigate the false positive rate (FPR), analyse the causes of these false positives, and the impact of the threshold on these. Next, we study the impact of our different heuristics from Section~\ref{sec:heuristics}, and the context sensitivity approach. We also discuss the patches we submitted. Finally, we evaluate the extensions we described earlier in Section~\ref{sec:extensions}.

\subsection{Evaluation of Past Security Vulnerabilities}
\label{pastevaluation}
First, we want to know whether \halo{} can indeed find race conditions in the Linux kernel with a clear security impact, and if so, at what detection threshold it does so. We investigated all CVEs for Linux from 2016 and 2017 that are related to missing locks and locking misuse. Table~\ref{tbl:pastvulns} lists the 14 vulnerabilities that are in scope.

We forward-ported these vulnerabilities to Linux v5.14.11 by reverting the patch. We evaluate on this modern kernel because only recently it has become possible to fully compile the Linux kernel under LLVM. Based on the fix patches, we identified the insufficiently-protected field access leading to the security issue\footnote{One issue, CVE-2016-9806, involves multiple fields being insufficiently-protected; but these all involve the same code, resulting in the same access counts and detection threshold. We hence grouped these into one row.}. We ran \halo{} on the vulnerable kernel, and checked whether this field access was indeed reported. Table~\ref{tbl:pastvulns} summarises these results, where we include the number of field accesses that are protected by a lock, the number of field accesses that are not protected by that lock, and the resulting minimal detection threshold. First, we see that 11 out of the 14 security vulnerabilities are indeed detected by \halo{}. Furthermore, 7 of these, i.e., half of the total number of vulnerabilities, are detected with a threshold less than 20\%. While most of these concern a single unlocked field access with multiple locked field accesses, some involve multiple unlocked accesses.

Three of these vulnerabilities stand out because they are not detected at all. The first, CVE-2016-9120 is a driver that has changed too much between 2016 and now to be able to forward-port the specific vulnerability, which we indicate with a \emph{?} in the table. The second, CVE-2017-6001, is more problematic. \halo{} cannot detect this issue because it also involves state becoming stale while waiting for the lock to be acquired. The solution thus not only involves changing the locking code, but also involves validating the acquired state~\cite{CVE-2017-6001}. We indicate this with a \emph{/} in the table. For CVE-2016-10200, there are 4 unlocked accesses versus only 1 locked. This issue involves a flags field that is checked for a security-sensitive flag using a helper function. It is furthermore dependent on the flag bit whether a lock is necessary or not. The safe functions heuristic results in our tool ignores the flags check and therefore this CVE cannot be detected with this heuristic enabled. We indicated cases that are filtered by heuristics as \emph{*}. If we were to disable this heuristic this leads to 16 unlocked accesses and 2 locked accesses.

The two remaining issues have a high threshold: CVE-2016-9806 and CVE-2017-12146. For the former, we have only one locked and one unlocked access. The latter case is an example of where multiple non-concurrent accesses artificially increase the required threshold.

\newcommand{\luntthreshold}[2]{\roundingtwodecimalplaces{100 * (#2 / (#1 + #2))}\%}
\newcommand{\cluntrow}[3]{ #1 & \luntthreshold{#2}{#3} & #2 & #3 }

\begin{figure}
\begin{center}
    \newcolumntype{t}{D{.}{.}{-3}}
	\begin{tabular}{  l | t | c | c }
		\multicolumn{2}{c}{} & \multicolumn{2}{c}{\textbf{\# of Field Accesses:}} \\
		\textbf{Vulnerability}  & \multicolumn{1}{c|}{\textbf{Threshold}} & \textbf{Locked} & \textbf{Unlocked} \\ 
		\hline \cluntrow{CVE-2017-15649   }{18}{ 1} \\
		\hline \cluntrow{CVE-2017-1000380 }{12}{ 1} \\
		\hline \cluntrow{CVE-2017-2636    }{ 8}{ 1} \\
		\hline \cluntrow{CVE-2016-2544    }{ 5}{ 1} \\
		\hline \cluntrow{CVE-2016-8655    }{25}{ 5} \\
		\hline \cluntrow{CVE-2017-7533    }{15}{ 3} \\
		\hline \cluntrow{CVE-2017-1000111 }{25}{ 5} \\
		\hline \cluntrow{CVE-2017-15265   }{ 3}{ 1} \\
		\hline \cluntrow{CVE-2016-7911    }{13}{ 7} \\
		\hline \cluntrow{CVE-2016-9806    }{ 1}{ 1} \\
		\hline \cluntrow{CVE-2017-12146   }{ 3}{ 3} \\
		\hline           CVE-2016-10200   & * & $\ast$ & $\ast$ \\
		\hline           CVE-2016-9120    & ? & ? & ? \\
		\hline           CVE-2017-6001    & / & / & / \\
	\end{tabular}
	\captionof{table}{Forward-ported vulnerabilities, with the minimal threshold required to detect them, and the number of locked \& unlocked accesses found.}\label{tbl:pastvulns} %
\end{center}
\end{figure}

\subsection{Evaluation of New Issues}

To evaluate the effectiveness of \halo{} to find \emph{new} race conditions, we manually studied new issues reported by our tool on Linux v5.14.11 in more detail.
To make this study tractable, we limited ourselves to the issues reported at (and below) a threshold of 16.67\%, initially \emph{with} context-sensitivity, and \emph{without heuristics and extensions}. We chose this threshold as half of the CVEs in Table~\ref{tbl:pastvulns} can be detected with it.
As we do not have access to a ground truth, we studied these issues and divided them into 3 main categories: probable true positive, probable false positive, and unknown. We add ``probable'' to emphasise the categorisation is according to our own judgement. We added an unknown category as there is code where we lacked the knowledge to fairly judge whether a report is a real bug (e.g., specific driver code). For clarity, in the remainder of this evaluation, we will no longer explicitly refer to true positives and false positives as being probable.

First, we investigated the reported issues without applying heuristics, in both the core kernel and the kernel modules---that is to say, we analyse all code that is compiled as modules for our default Debian config.
This resulted in \modulesandcorenoheuristicstotalissues{} potential issues: \modulesandcorenoheuristicstruepositives{} true positives, \modulesandcorenoheuristicsfalsepositives{} false positives, and \modulesandcorenoheuristicsunknown{} unknowns. If we ignore the unknowns---as it is unclear whether they are in fact true positives or false positives, they could go either way---this results in a FPR of \modulesandcorenoheuristicsfpr{}.
After applying our heuristics, this resulted in \modulesandcoreallheuristicstotalissues{} potential issues, of which most overlap with the initial reporting.
There were \modulesandcoreallheuristicstruepositives{} true positives, \modulesandcoreallheuristicsfalsepositives{} false positives, and \modulesandcoreallheuristicsunknown{} unknowns, or more concisely a FPR of \modulesandcoreallheuristicsfpr{}.
In total, we have categorised \totalpotentialissues{} unique potential issues, of which some, but not all, overlap between the reports with heuristics and without heuristics.

We also studied the impact of the context sensitivity. With only our heuristics enabled, and the context sensitivity disabled, we observe a false positive rate of 68.83\% (868 false positives and 393 true positives), up from 39.67\% where both context sensitivity and heuristics are enabled. Furthermore, disabling the context-sensitivity has a negative effect on the effectiveness of the heuristics. We study this effect in more detail in the next subsection.

Unfortunately, as is often the case for static analysers, many of the reported cases are false positives.
Despite this, the (potential) true positive rate is high enough that we easily found many new issues.
As we will discuss in more detail in Section~\ref{subsec:patches}, we created patches for a subset of the true positives reported at threshold 16.67\%. We limited ourselves to a subset where we felt most confident that the issue is real, and where the patch was fairly self-contained.

\subsection{Impact of Context and Heuristics}
\label{sec:causesoffalsepositivesandunknowns}
\begin{figure}
	\begin{center}
		\includegraphics[width=.5\textwidth]{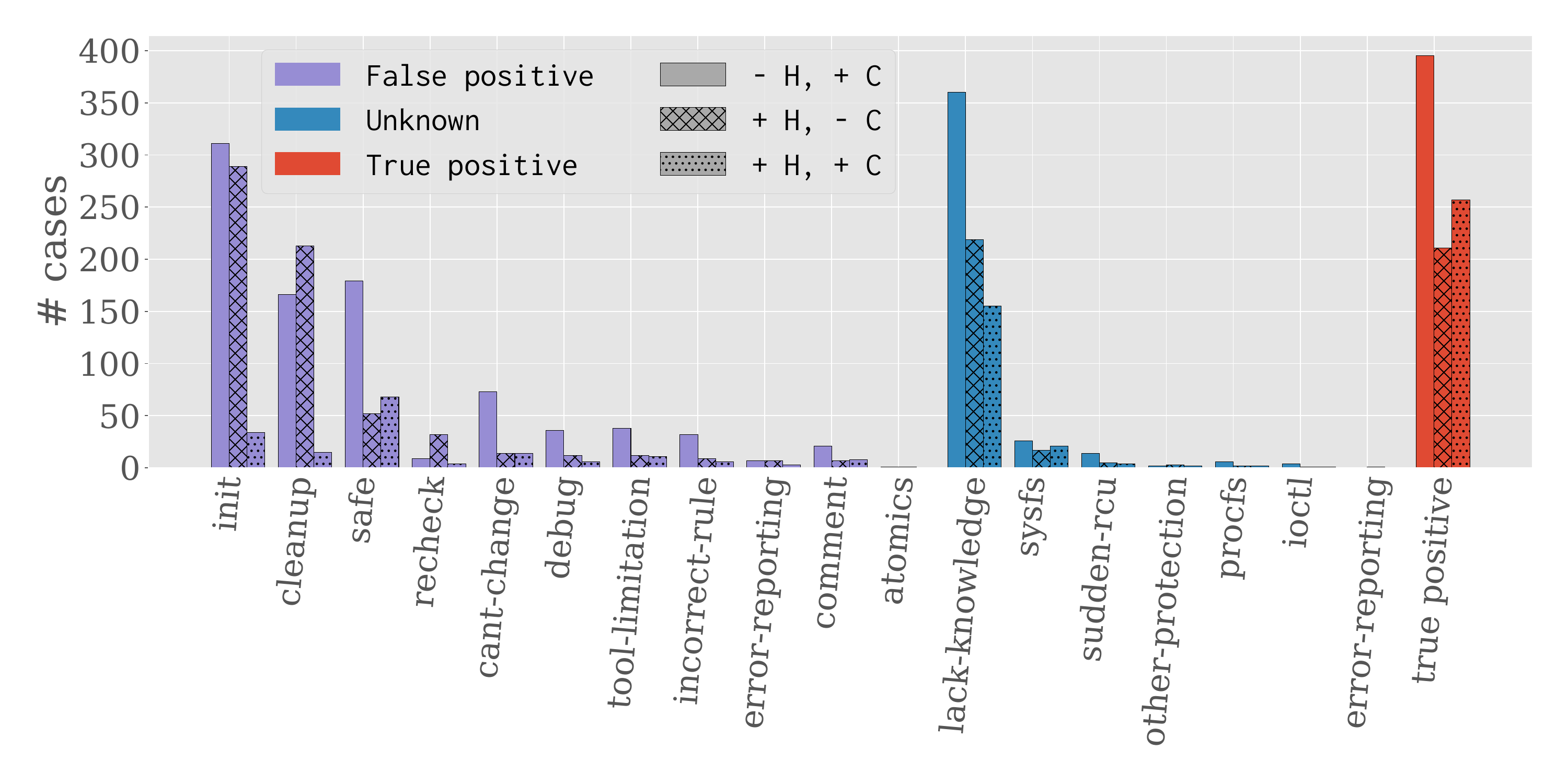}
	\end{center}
	\vspace{-4mm}
	\caption{Categorised cases at 16.67\% threshold, including a comparison of the influence of the heuristics and context-sensitivity. H indicates the heuristics, C indicates context-sensitivity. A plus symbol means the technique is enabled while a minus symbol means it is disabled.}
	\label{fig:evaluatedcases}
\end{figure}

We analysed the underlying causes of the false positive and unknown results in more detail, studying both the impact of our context-sensitivity and the heuristics. Figure~\ref{fig:evaluatedcases} shows the result of this analysis. The leftmost bars (i.e., the solid bars) show the results with no heuristics but with context-sensitivity enabled. The middle bars (i.e., the bars with a cross-hatch pattern) show the results with heuristics, but without context-sensitivity. Finally, rightmost bars (i.e., those with a dotted pattern) represent the result where context-sensitivity is combined with the heuristics. The bars are coloured according to their main category, i.e. purple means false positive, blue means unknown, and red means true positive.

The major cause of false positives is initialisation code, shown \textbf{init} in Figure~\ref{fig:evaluatedcases}. Interestingly, the difference between applying the heuristics and the context-sensitivity in isolation is relatively minor. This is because the cases pruned by the heuristics are from contexts where concurrency cannot happen, whereas the context-sensitive filtering takes into account the context and is thus able to deduce which contexts are \emph{not concurrent for the involved structs}. The two techniques are therefore complementary as they target different cases. Examples of non-concurrent contexts are device drivers that can be suspended and resumed, driver reset code, and driver IRQ code. This is hard to detect with our heuristics as these target generic initialisation patterns. This complementarity causes the combination of both context-sensitivity and heuristics to reduce these false positives significantly. The same occurs for the cleanup code, shown in the \textbf{cleanup} category.

Next, the \textbf{safe} category indicates that a lock is not needed for the reportedly violated field, for example because of coincidental other protections or benign data races (e.g., statistics, optimizations, ...). Even though we have a heuristic to detect this, context-sensitivity in isolation performs significantly better than this heuristic.

A related category is that where developers explicitly acknowledge in the code that a read or write might be racy, but that the resulting data race is benign. This might either be expressed in a comment above the access, or through the use of a call to \lstinline|data_race()|. We have categorised these cases as \textbf{comment} cases. One could ponder the validity of some of these comments when the comment provides no reason as to why the access is safe, but in all cases we have accepted these comments as-is. Benign races are a little more difficult to prune when using a context-sensitive technique when compared to using only heuristics because related locks can still be used in such benign contexts. Nevertheless, most of these cases are already eliminated by both the context-sensitivity as well as our heuristics. In particular, if comments indicating benign data races would be transformed into annotations such that we could completely automate this category of false positives, this would only reduce our FPR by \reducefprbyacceptingcomments{}.

Error messages often include values from fields. Therefore, these are usually false positives, except in the case that a pointer is dereferenced, then it can become tricky to reason about these cases, which is why we had to split this category into both a false positive and an unknown category for \textbf{error-reporting}. Debug code is closely related because fields in debug print statements are most often also not protected by a lock. This class of false positives can be eliminated by configuring the kernel to not include debug logging.

Using a lock is sometimes not necessary because protection is indirectly guaranteed by another component. An example is a flags field for drivers that ensures that some actions cannot happen concurrently. Our analyser is not aware of these relations and will therefore report violations of all these accesses. They are categorised as \textbf{other-protection-maybe}. Another example where protection is indirectly guaranteed is with atomic operations. These are categorised as \textbf{atomics}.

Another source of false-positive results are related to our tool's lack of precision in some areas. One of these involves the unlocked check--locked recheck paradigm, categorised as \textbf{recheck}. While the implementation of this heuristic indeed detects many instances of this paradigm (decreasing the FPR when it is enabled), our implementation is not able to detect \emph{all} such cases. Again, context-sensitivity also allows us to filter many of these cases automatically. Next, sometimes a field is accessed under an unrelated lock incidentally so many times that such fields fall below the 16.67\% threshold. These incorrect reports are categorised as \textbf{incorrect-rule}. Limitations in our tool's implementation (e.g., path-sensitivity) or in MLTA can also cause false positives. We categorise these as \textbf{tool-limitation}. Finally, there are instances of field accesses to variables that cannot change, which we categorise as \textbf{cant-change}. This is another example of pruning non-concurrent accesses. The context sensitivity infers there are no locks used that are related to the field.

For the unknown cases, \textbf{lack-knowledge} is our fallback category where we did not have enough knowledge to decide. While most of these are spread among many subsystems and drivers, some of these focus around locking in \textbf{ioctl}, \textbf{sysfs} and \textbf{procfs} code. Different implementations do not always follow the same locking rules for what are conceptually similar fields. Due to this causing uncertainty and our lack of knowledge of these systems, we have decided to categorise these in different unknown subcategories.
RCU protections can be used instead of locks, but they have to follow strict rules to be correct. However, sometimes code suddenly uses RCU protections instead of the previously-used locks. This \textit{could} be a bug, and we did find (and reported) issues related to this in the IPv6 subsystem. The validity of the remaining issues was unclear to us, which is why we included these under \textbf{sudden-rcu}.

When comparing the number of occurrences with and without heuristics enabled, it might appear counter-intuitive that some categories include \emph{more} reported issues when heuristics are enabled on top of the context-sensitivity, such as for example in the \textbf{safe} and \textbf{comment} categories, as well as in the \textbf{true positive} category. This is precisely \emph{because} the heuristics correctly filter out some unprotected field accesses, such as those seen in the reduction of the \textbf{init} and \textbf{cleanup}. Filtering out accesses with these heuristics can change the ratio of unlocked to total accesses, and thus bring certain cases below the threshold of our outlier-based approach. For example, consider a field with 2 out of 10 accesses being unlocked: 1 in the initialisation and 1 elsewhere. With a baseline run of \halo{} without heuristics, neither access is reported with a threshold of 16.67\%. However, upon filtering out the initialisations, the fraction becomes $\frac{1}{9} \leq 16.67\%$, which means that the remaining case \emph{will} be reported. This is a double-edged sword, as this can introduce both true positives and false positives. However, at least in our case, the number of added true positives outweighs the number of added false positives.

Although false positives take time to analyse, many are caused by easily recognisable categories, e.g. \textbf{init}, \textbf{cleanup}, and \textbf{safe}, they can be ignored very quickly. The practical impact of these false positives on analysists is therefore limited.

\subsection{Scalability of Threshold}
\begin{figure}
	\includegraphics[width=.5\textwidth]{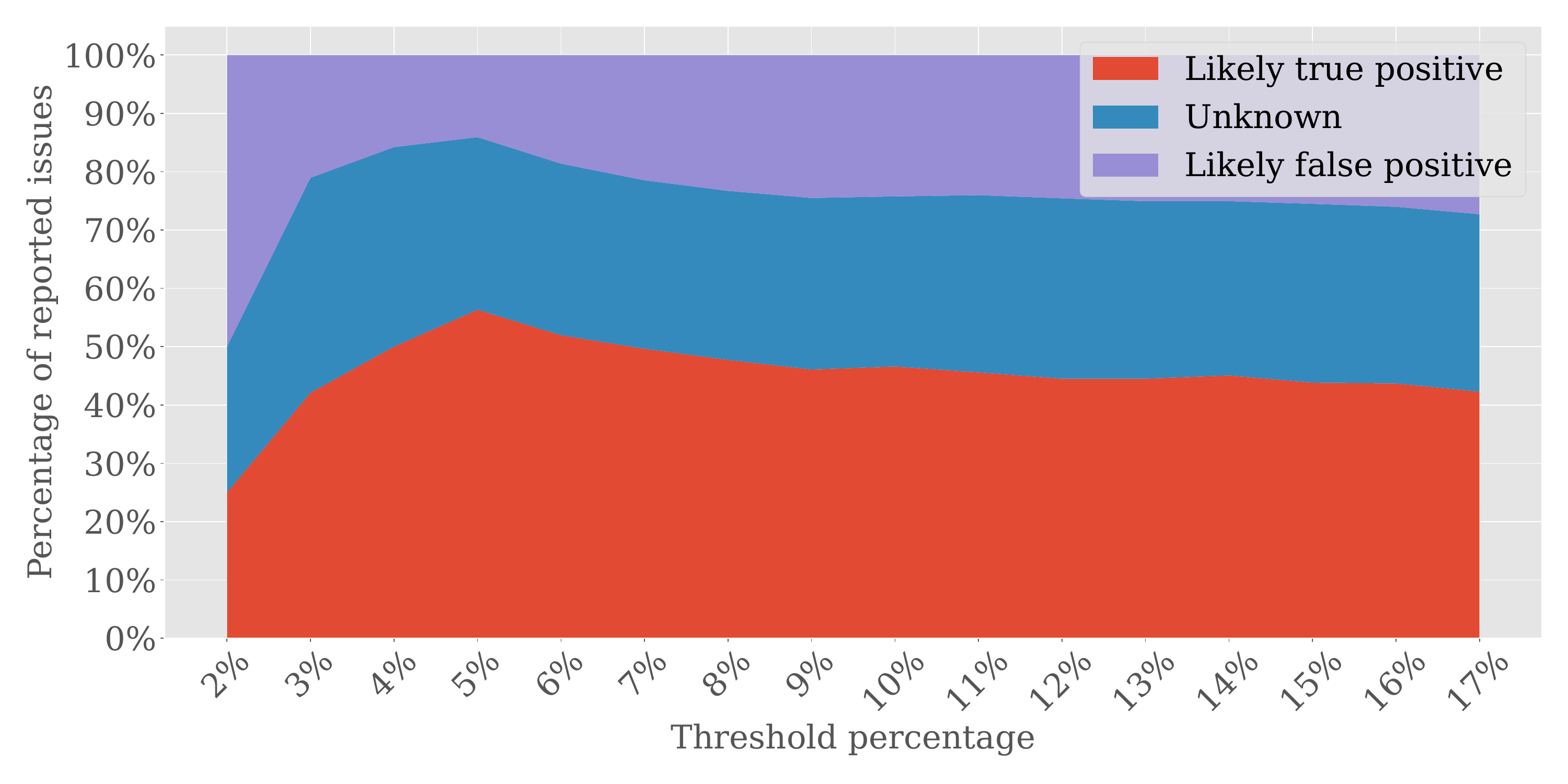}
	\vspace{-4mm}
	\caption{Division of reported issues in probable true positives, false positives, and unknowns (relative numbers)}
	\label{fig:headcategoriesnormalised}
\end{figure}

An important question to ask is how well the FPR scales with respect to the chosen threshold. As a lower threshold will include a subset of the issues reported with a higher threshold, we can study the evolution of the false positives and their causes for all thresholds up to and including 16.67\%. Figure~\ref{fig:headcategoriesnormalised} shows the results with relative counts. The division among the categories remains roughly the same with higher thresholds, but the absolute number of cases does increase.

\begin{figure}[t]
	\begin{center}
		\includegraphics[width=.5\textwidth]{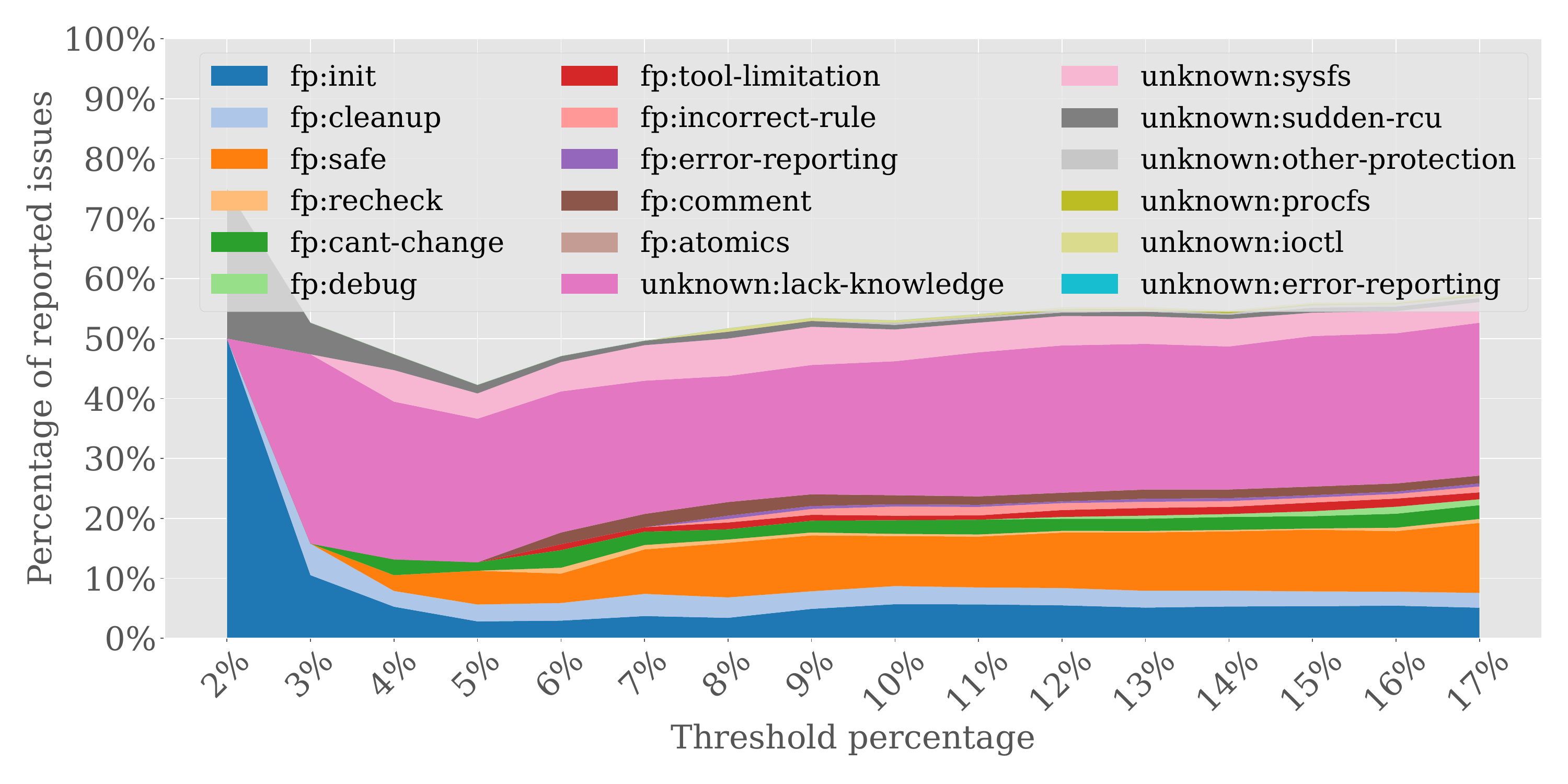}
	\end{center}
	\vspace{-4mm}
	\caption{Specific category breakdown, normalised}
	\label{fig:categoriesnormalised}
\end{figure}

We can also investigate the impact of the threshold on the specific categories of false positive and unknown. These results are shown in Figure~\ref{fig:categoriesnormalised}. We can see that \textbf{init} dominates the false positives at very low thresholds because these exceptional cases typically have many total accesses but only one initialisation. They quickly diminish in importance when the threshold is increased. Above a threshold of 7\%, the relative importance of most categories remains constant. Although a threshold of around 5\% would result in the lowest FPR, it would not result in many reported cases.

Finally, it is worth noting that very high values for the threshold will result in more cases where field accesses occur incidentally under a lock. In particular, the higher the threshold is, the higher the tolerance for counter-examples: above 50\% would treat cases with the majority of data uses unlocked and a minority of uses locked as violated.%

\subsection{Performance}

We have assessed the performance on an AMD Ryzen Threadripper 2990WX.

We start by evaluating the performance of \halo{} on our baseline Linux configuration \emph{excluding modules}, with a threshold of 16.67\%, running single-threaded. The run time was measured by taking the average of 10 runs. The total run time with \emph{no heuristics} is \runtimenomodulesnoheuristicssinglethreaded{}, and the total run time with \emph{all heuristics} active is \runtimenomodulesallheuristicssinglethreaded{}. When using 2 threads we measure a run time of \runtimenomodulesallheuristicstwothreads{}, but the run time stabilises around this time for higher thread counts.

We also evaluated the performance on Debian x86-64 with \emph{all modules included}, and \emph{all heuristics enabled}. With one thread we measured an execution time of \runtimemodulesallheuristicssinglethreaded{} for the Linux build with all modules included. When using 4 threads we measure a time of \runtimemodulesallheuristicsfourthreads{}, at which point the execution time stabilises when using more threads. There are two reasons for this stabilisation. First, loading the bitcode file takes about 3 minutes and uses only a single thread. Second, the generation of issue reports happens mostly sequentially. The run time impact of the context-sensitivity is negligible, upon disabling the context-sensitivity we obtained a run time of 8 minutes and 42 seconds down from 8 minutes 59 seconds.

We also evaluated the impact of the different heuristics on run time (RT) and on the FPR using a single thread. Each row in Table~\ref{tbl:heuristicsimpact} shows the result with \emph{only that heuristic} enabled.

It is clear that the reduction in FPR is not additive. Combinations of different heuristics eliminate false positive cases that were not eliminated by applying these heuristics separately (e.g. a case with an initialisation and a safe function call). It is interesting to note that most heuristics improve the run time because they reduce the amount of work to determine the code paths and source locations towards violations.

\begin{center}
	\resizebox{\columnwidth}{!}{
	\begin{tabular}{ | l | c | c |}
		\hline 
		\textbf{Heuristic} & \textbf{$\Delta$ FPR} & \textbf{$\Delta$ RT} \\ 
		\hline
		Initialisation and Clean-up & -13.49\% & -15.69\% \\
		\hline
		Unlocked Check--Locked Recheck & -1.76\% & +0.83\% \\ 
		\hline
		Write or Escape & -3.76\% & +1.81\% \\ 
		\hline
		Safe Functions & -8.28\% & -5.69\% \\ 
		\hline
		\textbf{Total} & -32.21\% & -19.86\% \\ 
		\hline
	\end{tabular}
	}
	\captionof{table}{Impact of heuristics on FPR and run time (RT)}\label{tbl:heuristicsimpact}
\end{center}

\subsection{Reported Issues and Submitted Patches}
\label{subsec:patches}
We also fixed some of the issues reported by \halo{} including its extensions. We started by verifying that the issue persists in more recent versions of Linux than 5.14.11 on which we ran \halo{}. As we did not wish to flood the Linux mailing lists with automatic reports without concrete fixes, we selected a subset of 32 issues that we deemed easy enough to fix ourselves. Appendix~\ref{sec:reportedissues}, and in particular Table~\ref{tbl:patches} describes our results in more detail.
To summarise the results, of the 32 submitted fixes, 21 have been accepted and committed to the mainline or linux-next kernel repository,
while an additional 2 have had the issue acknowledged by the subsystem maintainers who subsequently proposed a different fix. Of these, 15 fixes were backported to the appropriate stable kernel tree(s). Additionally, 1 patch has been reviewed and its issue was confirmed, but the patch has not yet been committed to a kernel repository. Of these 24 confirmed issues, we also investigated how long the issue has been
present in the kernel. The youngest bug was half a year old, while the oldest bug was 15 years old. The median age was 6.75 years and the average was 5.61 years.

Note that of these 32 issues we investigated, fixed, and reported, 2 were rejected by the kernel developers as false positives. The first issue
concerns the \lstinline|iommu_page_response| function: a security check on a field and the actions that acted upon that field did not happen
atomically due to the lack of a lock. However, it turns out that the issue is not exploitable in practice because the actions
required to exploit this are only possible if an IOMMU driver would contain a bug. However, the maintainers agreed the code looks strange,
and proposed to improve the documentation of the involved functions. %
The second issue concerns the \lstinline|tipc_node_timeout| function, which explicitly takes a \emph{read lock} on a struct, but then only \emph{writes} to that locked struct.
According to the maintainers, the code is structured in such a way that there cannot be concurrent updates,
and as such writing is safe~\cite{lore_tipc_reply_blinded}. %

Furthermore, we first considered some issues as unknown cases, but later as probable true positives. This was the case for statfs functions. The locking is often inconsistent in these functions, such as the implementation in the f2fs file system, which performs locking for \textit{some} fields of the super block, but not all of them. In other cases such as the shmem's implementation, no locking happens at all. The f2fs maintainers confirmed that the f2fs issue is a real one, because it could lead to unreliable, inconsistent results. Therefore, we have categorised all statfs cases as probable true positives.

\subsection{Extensions}
Finally, we report on the effectiveness of our extensions. The Lockdep extension reports 4 true positives. Although all of these were already reported by \halo{} without this extension, it allows us to focus on this specific type of issue without false positive results. The extension to find read-write lock violations reports 3 results: two true positives (one of which was acknowledged by the maintainers, and one of which we have not received a reply yet), and one false positive issue---this is the \lstinline|tipc_node_timeout| issue we mentioned above.

\section{Related Work}
\label{sec:related}
Many approaches exist to analyse complex software systems.
We focus on some relevant state-of-the-art techniques to analyse kernel and low-level
code, and highlight similarities and differences to our own work. First, we compare our work against other approaches to analyse locks in kernels. Then, we compare our work against other outlier-based analyses.

\subsection{Locking Bug Analysers}
There are multiple ways to compare and contrast analyses for locking bugs. A first categorisation is static versus dynamic techniques. Static analysers analyse code bases without executing them. As an advantage, they can consider many architectural configurations and devices, and do not need to find specific inputs that trigger a potential issue. As a downside, they are potentially not very precise, as these tools need to reason over complex interactions in large code bases. Our tool is a static analysis tailored for complex systems code.

Static analysis can start from annotations. For example, Sparse~\cite{sparse} is a tool tailored to Linux which can detect missing locks, based on manual annotations indicating which locks should be held, acquired, or released by a function. Its accuracy and usefulness is thus limited to code with these annotations. %
Static techniques do not need to depend on such annotations. For example, RacerX~\cite{racerx} utilises a flow-sensitive, inter-procedural analysis to find both race conditions and deadlocks. It is based on lockset analysis, like most static techniques to detect concurrency bugs~\cite{dcuaf,dlos,deligiannis2015fast,racerx,kahlon2009static,goblint,relay}. Lockset analysis can also be used in deadlock detection~\cite{dlos,dsac,racerx}. Some techniques furthermore use symbolic locksets or symbolic execution~\cite{deligiannis2015fast,goblint,relay}.
 
\begin{figure}
\begin{center}
    \newcolumntype{t}{D{.}{.}{-3}}
    \resizebox{\columnwidth}{!}{
    \begin{tabular}{ | l | c | c | c | c | }
        \hline
        \textbf{Tool}  & \textbf{Class} & \textbf{Goal} & \textbf{Target} & \textbf{Indirect calls} \\ 
        \hline RacerX~\cite{racerx} & LS & RC, DL & \fullcirc[1ex] & None \\
        \hline Goblint~\cite{goblint} & SLS & RC & \halfcirc[1ex] & None \\
        \hline DLOS~\cite{dlos} & LS & DL & \fullcirc[1ex] & None \\
        \hline SDILP~\cite{sdilp} & LS & RC & \halfcirc[1ex] & None \\
        \hline DSAC~\cite{dsac} & C & SAC & \threefourthscirc[1ex] & Type-based \\
        \hline DCUAF~\cite{dcuaf} & LS & UAF RC & \halfcirc[1ex] & Type-based \\
        \hline RELAY~\cite{relay} & RLS & RC & \fullcirc[1ex] & None \\
        \hline WHOOP~\cite{deligiannis2015fast} & SLS & RC & \halfcirc[1ex] & None \\
        \hline \textbf{\halo{}} & O & RC & \fullcirc[1ex] & Type-based \\
        \hline
    \end{tabular}
    }
    \captionof{table}{
        Comparison with related static and hybrid analysers for locking bugs in kernels.
        The RC, DL, SAC and UAF in the Goal field represent race condition, deadlock, sleep-in-atomic context and use-after-free.
        LS means lockset, SLS means symbolic lockset, RLS means relative lockset, C means custom and O means outlier-based.
        \fullcircnoshift[1ex] indicates the whole kernel is analysed, \threefourthscircnoshift[1ex] indicates device drivers and filesystems are analysed and \halfcircnoshift[1ex] indicates only device drivers are analysed.
    }\label{tbl:positioning}
\end{center}
\end{figure}

Table~\ref{tbl:positioning} shows a comparison of our tool with the most relevant static and hybrid locking analysers of kernels. RacerX~\cite{racerx} is the original tool for static race detection in kernels. Goblint~\cite{goblint} improves upon RacerX by representing variables symbolically, and by using alias analysis to create equivalence classes of variables that may alias. Rather than relying on structs (such as RacerX), or on field accesses (such as our approach), it depends only on whether pointers alias. Goblint is limited to detecting race conditions inside device drivers. %
RELAY~\cite{relay} also uses symbolic execution and a variant of lockset analysis called relative lockset analysis to make its analysis more scalable. The authors have also introduced multiple heuristics to filter probable false positives. Unfortunately, their technique takes a long time to run because of the symbolic execution, and cannot analyse the majority of kernel functions. WHOOP~\cite{deligiannis2015fast} is a symbolic approach that uses symbolic locksets to find races in drivers.
Static race detectors typically suffer from a high false positive rate because they assume all functions can be executed concurrently. DCUAF~\cite{dcuaf} recognises this and derives which kernel interfaces can execute concurrently. However, it focusses on one specific type of concurrency bug: concurrency use-after-frees. These two key ideas allow DCUAF to achieve a very low false positive rate.
DLOS~\cite{dlos} uses a path-sensitive approach to detects deadlocks in kernels, that improves upon RacerX~\cite{racerx} by not considering all paths to be able to concurrently execute. DSAC~\cite{dsac} finds uses of locking APIs that are illegal in certain contexts. Finally, SDILP~\cite{sdilp} is a hybrid approach that first uses dynamic analysis to find potential races and then checks for locally similar patterns statically to find races that were not triggered during execution.

Not all static race detectors focus on kernel code~\cite{blackshear2018racerd,kahlon2009static}. RacerD~\cite{blackshear2018racerd} uses abstract interpretation to reason about data races, but is tailored to Java code.

Dynamic analyses execute code as part of their analysis step. This indicates their major downside: the faulty code needs to be executed, which might not always be trivial to achieve. Furthermore, while they suffer less from issues regarding the precision of underlying analyses, they do not eliminate false positives caused by benign races, just like most analyses. Dynamic analyses for finding race conditions in the Linux kernel include Lockdep~\cite{lockdep}, which detects lock ordering issues and missing locks by using assertions; Kernel Concurrency Sanitizer (KCSAN)~\cite{lwn_concurrency_part1}, which is based on the approach of DataCollider~\cite{DataCollider} to randomly sample a fraction of the memory accesses using breakpoints; Eraser~\cite{eraser}, which was the first approach to introduce lockset analysis. Race conditions can also be found through fuzzing, as shown by Razzer~\cite{razzer} and Conzzer~\cite{conzzer}. Dynamic techniques can also be used to infer locking rules: LockDoc~\cite{lockdoc} creates documentation of locking rules based on outlier-based analysis of memory traces.

Finally, hybrid techniques~\cite{sdilp,chen2019detecting} have been used successfully in the past as well.

\subsection{Outlier-Based Analysers}
Static outlier-based techniques have been employed to find security issues. They all rely on the underlying idea that most code is correct, and deviations from patterns can signal a bug. This idea was first introduced by Engler et al. in ``Bugs as Deviant Behavior''~\cite{engler2001bugs}. PeX~\cite{zhang2019pex} identifies mappings between permission checks and security-sensitive functions to detect missing permission checks. LRSan~\cite{lrsan} targets finding TOC-TOU bugs, and CHEQ~\cite{cheq} targets finding NULL-pointer dereferences, missing error handling, and double-fetch bugs.
Crix~\cite{crix} improves upon LRSan and targets detecting missing security checks. %
Finally, APISan~\cite{apisan} derives rules about API usage and checks for violations against them, allowing it to detect missing unlock calls.

\section{Conclusion}
\label{sec:conclusion}
We introduced a novel static analysis technique to find potential race conditions in complex system software.
We enhanced the outlier-based aspect of the technique with a scalable context-sensitive mechanism.
We implemented this technique in \halo{}, and have successfully applied it on Linux, where we found new bugs. We have evaluated \halo{}: its ability to find existing CVEs, and its ability to find new issues.  Furthermore, we studied the reasons for false positive results, and the impact that parameters, context-sensitivity, heuristics, and extensions have on the results. In addition, we submitted patches to fix some reported issues, of which 24 have been confirmed.

\section*{Availability}

We will release our code publicly on GitHub, and will submit to the Artefact Evaluation.%

\appendix
\section{Confirmed Issues}
\label{sec:reportedissues}
\begin{center}
	\resizebox{\columnwidth}{!}{
	\begin{tabular}{ | l | c | c | }
		\hline
		\textbf{File(s)} & \textbf{Accepted} & \textbf{Impact} \\ 
		\hline
		drivers/firewire/core-card.c & \checkmark & memory corruption \\   
		\hline
		drivers/infiniband/.../qp.c * 2 & \checkmark & memory corruption  \\  
		\hline
		drivers/.../alx/main.c & \checkmark & crash \\  
		\hline
		drivers/.../sfc/mcdi.c & \checkmark & data corruption \\  
		\hline
		\makecell{drivers/nvme/target/\\admin-cmd.c,configfs.c,\\core.c,nvmet.h,zns.c} & \makecell{Fixed by\\maintainer} & memory corruption  \\ 
		\hline
		drivers/.../ath11k/mac.c & \checkmark & memory corruption \\  
		\hline
		drivers/.../mwifiex/11h.c & \checkmark & memory corruption \\  
		\hline
		drivers/usb/usbip/stub\_rx.c & \checkmark &  inconsistent configuration \\  
		\hline
		fs/btrfs/space-info.c & \checkmark & performance degradation \\  
		\hline
		fs/ceph/caps.c & \checkmark & data corruption  \\  
		\hline
		fs/dcache.c &  & broken fail-safe \\  
		\hline
		fs/f2fs/super.c & \checkmark & inconsistent data \\  
		\hline
		net/bluetooth/hci\_conn.c & \checkmark & memory corruption \\  
		\hline
		net/bluetooth/hci\_event.c & \checkmark & crash \\  
		\hline
		net/../hci\_event.c * 3 & \checkmark & memory corruption \\  
		\hline
		net/bluetooth/hci\_request.c & \checkmark & crash \\  
		\hline
		net/bluetooth/mgmt.c & \checkmark & event never fires \\  
		\hline
		net/core/devlink.c & \makecell{Fixed by\\maintainer} & memory corruption \\  
		\hline
		net/ipv6/addrconf.c & \checkmark & inconsistent configuration \\  
		\hline
		net/ipv6/addrconf.c * 2 & \checkmark & memory corruption \\
		\hline
    \end{tabular}}
	\captionof{table}{List of the patches for issues that have been confirmed by the kernel maintainers, their status, and their impact}
	\label{tbl:patches}
\end{center}
Table~\ref{tbl:patches} describes the status of the \emph{confirmed issues} and patches. Multiple rows for the same file(s) mean that there were multiple different bugs in the same file; a \textit{* N} means that a single patch fixed $N$ issues. Accepted means the patch has been committed to either the mainline (i.e. the current release of Linux) or linux-next repository.

\bibliographystyle{plain}
\bibliography{bibliodatabase}

\end{document}